\documentclass[twocolumn,showpacs,preprintnumbers,amsmath,amssymb]{revtex4}
\usepackage{amsmath}
\usepackage{amssymb}
\usepackage{graphicx}
\usepackage{bm}
\usepackage{epstopdf}
\usepackage{slashed}

\begin{document}

\title {Field sources in a Lorentz symmetry breaking scenario with a single background vector}

\author{L.H.C. Borges}
\email{luizhenriqueunifei@yahoo.com.br}
\affiliation{Universidade Federal do ABC, Centro de Ci\^encias Naturais e Humanas, Rua Santa Ad\'elia, 166, 09210-170, Santo Andr\'e, SP, Brasil}

\author{F.A. Barone}
\email{fbarone@unifei.edu.br}
\affiliation{IFQ - Universidade Federal de Itajub\'a, Av. BPS 1303, Pinheirinho, Caixa Postal 50, 37500-903, Itajub\'a, MG, Brazil}

\author{J.A. Helay\"el-Neto}
\email{helayel@cbpf.br}
\affiliation{Centro Brasileiro de Pesquisas F\'\i sicas, Rua Dr.\ Xavier Sigaud 150, Urca, 22290-180, Rio de Janeiro, RJ, Brazil}

\begin{abstract}
This paper is devoted to investigating the interactions between stationary sources of the electromagnetic field, in a model which exhibits explicit  Lorentz-symmetry breaking due to the presence of a single background vector. We focus on physical phenomena that emerge from this kind of breaking and which have no counterpart in Maxwell Electrodynamics.     
\end{abstract}

\maketitle

\section{Introduction}
\label{I}

Since the proposal of Carroll, Field and Jackiw in \cite{CFJ}, a great deal of efforts have been devoted to the study of models which exhibit explicit Lorentz-symmetry braking in many contexts. Among them, stand out models where the Lorentz-symmetry is broken in the gauge field sector. We can mention, for instance, the study of classical electrodynamics with Lorentz symmetry violation \cite{PRDqbk2004,EJPCxw2006,PRDbfho2003,PRDlp2004,PRLa2007}, electromagnetic wave propagation \cite{PRDtwfh2005,ABBGHplb2012,CFMepjc2012,ra1,ra2,NPBks2011,PRDggcm2010}, QED effects \cite{PLBck2001,PRDhlpw2009,NPBks2011}, the extension of the Standard Model for Lorentz symmetry breaking scenarios \cite{PRDck1997,PRDck1998}, non-minimal couplings \cite{EPJCbccho2009,PRDbsfo2011,GFSSNjpg2012}, extensions of Lorentz-symmetry breaking with derivatives of higher order \cite{PRDkm2009}, the Casimir effect \cite{PRDft2006}, classical point-like particles dynamics \cite{PLBkr2010} and so on.

Regarding the Abelian gauge fields, we have two kinds of models which exhibit Lorentz-symmetry breaking: the tensorial model \cite{PRDqbk2004} and the Carroll-Field-Jackiw one \cite{CFJ}, which is a generalization of the Chern-Simons model for $(3+1)$ dimensions. The first one is richer in comparison with the second one, once it has more parameters and, so, can lead to more physical effects with no counterpart in Lorentz invariant theories. 

One of the most fundamental questions that can be done in any gauge theory is about the interaction between external field sources. This subject has been considered in the context of Lorentz-symmetry breaking theories \cite{PRDqbk2004}, but some points are still worth understanding, not only for their theoretical aspects, but also because of their possible experimental relevance in the search for physical phenomena with no counterparts in Maxwell theory. 

This paper is devoted to investigating  electromagnetic phenomena in a Lorentz symmetry breaking scenario, with no counterpart in Maxwell theory.  We consider a tensorial model  and restrict our analisys to the non birefringent sector \cite{PRDqbk2004}, parameterizing the background tensor in such a way to reduce it to one background vector, what allows us to obtain exact results. We also focus our attention on the case of stationary sources. 

Specifically, we obtain exactly the photon propagator of the model in Section (\ref{II}). In section (\ref{III}) we consider effects due to the presence of point-like stationary charges. In Section (\ref{IV}) we compute the interaction energy between a steady line current and a point-like stationary charge. Section (\ref{V}) is devoted to the study of physical phenomena due to the presence of one or two Dirac strings, and section (\ref{conclusoes}) is devoted to our final remarks and conclusions.

\section{The model}
\label{II}

In this paper let us consider the electromagnetic field in a Lorentz-symmetry breaking scenario parametrized by just one background vector. The Lorentz-symmetry breaking is accomplished by a tensorial coupling of the electromagnetic field in the Lagrangian, as follows:
\begin{eqnarray}
\label{lagEm}
{\cal L}=-\frac{1}{4}F_{\mu\nu}F^{\mu\nu}-\frac{1}{2\gamma}\left(\partial_{\mu}A^{\mu}\right)^{2}-\frac{1}{2}v^ {\mu}v_{\nu}F_{\mu\lambda}F^{\nu\lambda}\cr
+J^{\mu}A_{\mu}\ ,
\end{eqnarray}
where $A^{\mu}$ is the electromagnetic field, $F^{\mu\nu}=\partial^{\mu}A^{\nu}-\partial^{\nu}A^{\mu}$ is the field strength, $J^{\mu}$ is the external source, $\gamma$ is a gauge parameter and $v^\mu$ is the background vector. We will be working in a $3+1$ dimensional space-time with Minkowsky metric $\eta^{\mu\nu}=(+,-,-,-)$. 

Once the Lorentz-symmetry breaking must be very tiny, the vector $v^{\mu}$, which is a dimensionless quantity, must be small.

In reference \cite{GFSSNjpg2012}, two points are discussed in connection to the model (\ref{lagEm}). The first one is the fact that the Lagrangian (\ref{lagEm}) can be obtained from a model with a non-minimal coupling of the electromagnetic field with the spinorial one. The second point relies on the fact that (\ref{lagEm}) is a special case of the Lagrangian proposed in \cite{PRDqbk2004}, where we have a tensorial parameter controlling the Lorentz symmetry breaking with a term $\sim k_{\mu\nu\alpha\beta}F^{\alpha\beta}F^{\mu\nu}$. That is, when $k_{\mu\nu\alpha\beta}\sim\eta_{\mu\alpha}v_{\nu}v_{\beta}-\eta_{\nu\alpha}v_{\mu}v_{\beta}+\eta_{\nu\beta}v_{\mu}v_{\alpha}-\eta_{\mu\beta}v_{\nu}v_{\alpha}$ the model discussed in \cite{PRDqbk2004} is given by (\ref{lagEm}). 

The propagator $D^{\mu\nu}(x,y)$ satisfies the differential equation
\begin{eqnarray}
\label{EDProEm}
\Biggl[\left(\partial^{2}+(v\cdot\partial)^{2}\right)\eta^{\mu\nu}-\left(1-\frac{1}{\gamma}\right)\partial^{\mu}\partial^{\nu}+v^{\mu}v^{\nu}\partial^{2}\nonumber\\
-\left(v\cdot\partial\right)v^{\mu}\partial^{\nu}-\left(v\cdot\partial\right)v^{\nu}\partial^{\mu}\Biggr]{D_{\nu}}^{\lambda}\left(x,y\right)=\nonumber\\
=\eta^{\mu\lambda}\delta^{d+D+1}\left(x-y\right) \ ,
\end{eqnarray}
where $\partial^{2}=\partial^{\mu}\partial_{\mu}$ and $v\cdot\partial=v^{\mu}\partial_{\mu}$.

Finding out the Fourier transform and taking the Lorentz gauge condition, $\gamma=1$, one can readily show that the propagator is given by 
\begin{eqnarray}
\label{propEm}
D^{\mu\nu}(x,y)=-\int\frac{d^{4}p}{(2\pi)^{4}}\frac{e^{-ip\cdot(x-y)}}{[p^{2}+(p\cdot v)^2]}\Biggl[\eta^{\mu\nu}-\frac{v^{\mu}v^{\nu}}{1+v^{2}}\cr\cr
-\frac{v^{2}}{1+v^{2}}\frac{(p\cdot v)^2}{p^{2}}\frac{p^{\mu}p^{\nu}}{p^{2}}+\frac{1}{1+v^{2}}\frac{(p\cdot v)}{p^{2}}(p^{\mu}v^{\nu}+v^{\mu}p^{\nu})\Biggr] \ .\nonumber\\
\ 
\end{eqnarray}

\section{Point-like charges}
\label{III}

In this section we study the interaction energy between two stationary point-like field sources in $(3+1)$ dimensions. The results can be easily extended to the an arbitrary number of spatial dimensions.

In the first model, we consider the field sources described by the time-independent current
\begin{eqnarray}
\label{corre1Em}
J_{\mu}^{I}({\bf x})=\sigma_{1}\eta^{0}_{\mu}\delta^{3}\left({\bf x}-{\bf a}_ {1}\right)+\sigma_{2}\eta^{0}_{\mu}\delta^{3}\left({\bf x}-{\bf a}_ {2}\right) \ ,
\end{eqnarray}
where we have two spatial Dirac delta functions, concentrated at the positions ${\bf a}_{1}$ and ${\bf a}_{2}$. The parameters $\sigma_{1}$ and $\sigma_{2}$ are the coupling constants between the field and the delta functions and can be seen as electric charges. 

Once we have a quadratic Lagrangian in the field variables, as discussed in references \cite{Zee,BaroneHidalgo1,BaroneHidalgo2}, the contribution due to the sources to the ground state energy of the system is given by
\begin{equation}
\label{zxc1}
E=\frac{1}{2T}\int\int d^{4}x\ d^{4}y J^{\mu}(x)D_{\mu\nu}(x,y)J^{\nu}(x)
\end{equation}
where $T$ is the time variable.

Substituting (\ref{propEm}) and (\ref{corre1Em}) in (\ref{zxc1}), discarding the self-interacting contributions (the interactions of a given point-charge with itself), computing the integrals in the following order,  $d^{3}{\bf x}$, $d^{3}{\bf y}$, $dx^{0}$, $dp^{0}$ and $dy^{0}$, using the Fourier representation for the Dirac delta function $\delta(p^{0})=\int dx/(2\pi)\exp(-ipx)$ and identifying the time interval as $T=\int dy^{0}$, we can write
\begin{eqnarray}
\label{Ener2EM}
E^{I}=\sigma_{1}\sigma_{2}\left(1-\frac{(v^{0})^{2}}{1+v^{2}}\right)\int\frac{d^{3}{\bf p}}{(2\pi)^{3}}\frac{\exp(i{\bf p}\cdot{\bf a})}{[{\bf p}^2-({\bf v}\cdot{\bf p})^{2}]} \ ,
\end{eqnarray}
where we defined $\bf{a}={\bf a}_{1}-{\bf a}_{2}$, which is the distance between the two electric charges.

In order to calculate the integral in (\ref{Ener2EM}), we first split the vector $\bf p$ into two parts, one parallel, ${\bf p}_{p}$, and the other normal, ${\bf p}_{n}$, to the vector $\bf v$, namely
\begin{eqnarray}
\label{mudan1EM}
{\bf p}={\bf p}_{n}+{\bf p}_{p},\ \ {\bf p}_{p}={\bf v}\Bigl(\frac{{\bf v}\cdot{\bf p}}{{\bf v}^{2}}\Bigr),\ \ {\bf p}_{n}={\bf p}-{\bf v}\Bigl(\frac{{\bf v}\cdot{\bf p}}{{\bf v}^{2}}\Bigr),\cr
\ 
\end{eqnarray}
where ${\bf p}_{n}\cdot{\bf v}=0$. Now we define the vector ${\bf q}$
\begin{eqnarray}
\label{defq}
{\bf q}={\bf p}_{n}+{\bf p}_{p}\sqrt{1-{\bf v}^{2}}={\bf p}+{\bf v}\Bigl(\frac{{\bf v}\cdot{\bf p}}{{\bf v}^{2}}\Bigr)(\sqrt{1-{\bf v}^{2}}-1)\ .
\end{eqnarray}

With definitions (\ref{mudan1EM}) and (\ref{defq}) one may write
\begin{eqnarray}
\label{mudan6EM}
{\bf p}_{p}=\frac{\bf v(\bf v\cdot\bf q)}{{\bf v}^{2}\sqrt{1-{\bf v}^{2}}}\ \ \ ,\ \ \ {\bf p}_{n}=\bf q-\frac{\bf v(\bf v\cdot\bf q)}{{\bf v}^{2}}\cr\cr
\Rightarrow {\bf p}={\bf q}+\frac{(\bf{v}\cdot{\bf q})\bf{v}}{{\bf v}^{2}}\left(\frac{1}{\sqrt{1-{\bf v}^{2}}}-1\right)  \ ,
\end{eqnarray}
\begin{equation}
\label{zxc2}
{\bf q}^{2}={\bf p}^2-({\bf v}\cdot{\bf p})^{2}
\end{equation}

With the aid of the definition
\begin{equation}
\label{zxc3}
{\bf b}={\bf a}+\frac{1-\sqrt{1-{\bf v}^{2}}}{\sqrt{1-{\bf v}^{2}}}\frac{{\bf v}\cdot{\bf a}}{{\bf v}^{2}}{\bf v}\ ,
\end{equation}
and (\ref{mudan6EM}), we also have
\begin{eqnarray}
\label{mudan3EM}
{\bf p}\cdot{\bf a}={\bf b}\cdot{\bf q} 
\end{eqnarray}

The Jacobian of the transformation from ${\bf p}$ to ${\bf q}$ can be calculated from (\ref{mudan6EM})
\begin{equation}
\label{mudan5EM}
\det\left[\frac{\partial\bf{p}}{\partial\bf{q}}\right]=\frac{1}{\sqrt{1-{\bf v}^{2}}} \ .
\end{equation}

Substituting (\ref{zxc2}) and (\ref{mudan3EM}), in (\ref{Ener2EM}) and using eq. (\ref{mudan5EM}) we have
\begin{eqnarray}
\label{Ener3EM}
E^{I}=\frac{\sigma_{1}\sigma_{2}}{\sqrt{1-{\bf v}^{2}}}\left(1-\frac{(v^{0})^{2}}{1+v^{2}}\right)\int\frac{d^{3}{\bf q}}{(2\pi)^{3}}\frac{\exp(i{\bf b}\cdot{\bf q})}{{\bf q}^2} \ .
\end{eqnarray}

Using the fact that \cite{BaroneHidalgo1} 
\begin{eqnarray}
\label{Ener4EM}
\int\frac{d^{3}{\bf q}}{(2\pi)^{3}}\frac{\exp(i{\bf b}\cdot{\bf q})}{{\bf q}^2}=\frac{1}{4\pi|{\bf b}|} \ ,
\end{eqnarray}
as well as eq. (\ref{zxc3}), and performing some simple manipulations we have, finally,
\begin{eqnarray}
\label{Ener6EM}
E^{I}=\frac{\sigma_{1}\sigma_{2}}{4\pi}\frac{\sqrt{1-{\bf v}^{2}}}{1+v^2}\left[{\bf a}^{2}+\frac{({\bf v}\cdot{\bf a})^{2}}{1-{\bf v}^{2}}\right]^{-1/2} \ .
\end{eqnarray}

Equation (\ref{Ener6EM}) is an exact result and gives the interaction energy between two point-like charges for the model (\ref{lagEm}). If we take $v^{\mu}=0$, expression (\ref{Ener6EM}) reduces to the well-known Coulombian interaction. In (\ref{Ener6EM}) the coefficient $\sqrt{1-{\bf v}^{2}}/(1+v^2)$ can be absorbed into the definition of the electric charges $\sigma_{1}$ and $\sigma_{2}$ and does not indicates a Lorentz symmetry braking. The extra factor proportional to $({\bf v}\cdot{\bf a})^{2}$ inside the brackets is an evident contribution which evinces the Lorentz symmetry breaking. It leads to an anisotropic interaction between the charges. This fact can be clarified if we compute the force on the charge $2$, for instance,
\begin{equation}
\label{for1}
{\bf F}^{I}=-\nabla E^{I}=\frac{\sigma_{1}\sigma_{2}}{4\pi}\frac{1-{\bf v}^{2}}{1+v^{2}}\frac{1}{{\bf a}^{2}}\frac{(1-{\bf v}^{2}){\hat a}+({\bf v}\cdot{\hat a}){\bf v}}{\left[1-{\bf v}^{2}+({\bf v}\cdot{\hat a})^{2}\right]^{3/2}}\ .
\end{equation}

Notice that (\ref{for1}) is a long range but anisotropic interaction. In the special situation where ${\bf v}$ and ${\hat a}$ are perpendicular each other, the force (\ref{for1}) becomes a Coulombian interaction with effective charges $\sigma\to\sigma\sqrt{\frac{(1-{\bf v}^{2})^{1/2}}{1+v^{2}}}$. In the special case where ${\bf v}=0$ the force (\ref{for1}) still exhibits a Coulombian-like behavior, with electric charges renormalized by $\sigma\to\sigma[1+(v^{0})^{2}]^{-1/2}$.

Once $v^{\mu}$ is a small quantity, it is relevant to expand expression (\ref{for1}) in lowest order in $v^{\mu}$,
\begin{eqnarray}
\label{for1exp}
{\bf F}^{I}\cong\frac{\sigma_{1}\sigma_{2}}{4\pi}\frac{1}{{\bf a}^{2}}\Biggl[\Biggl(1-(v^{0})^{2}+\frac{1}{2}{\bf v}^{2}\Biggr){\hat a}\nonumber\\
+({\bf v}\cdot{\hat a})\Biggl({\bf v}-\frac{3}{2}({\bf v}\cdot{\hat a}){\hat a}\Biggr)\Biggr]\ .
\end{eqnarray}
The first contribution inside the brackets in (\ref{for1exp}), proportional to ${\hat a}$, exhibits a typical inverse-square radial law. The second contribution inside the brackets is a long-range one but exhibits anisotropy.

An important consequence of the anisotropies in expression (\ref{Ener6EM}) is the emergence of an spontaneous torque on an electric dipole. In order to investigate this effect, let us consider a typical dipole composed by two opposite electric charges placed at a fixed distance apart. For this task we use expression (\ref{for1}) with $\sigma_{1}=-\sigma_{2}=\sigma$, ${\bf a}_{1}={\bf R}+\frac{{\bf A}}{2}$ and ${\bf a}_{2}={\bf R}-\frac{{\bf A}}{2}$, where we take the distance vector $\bf A$ fixed (and small). Thus the interaction energy between two charges (\ref{Ener6EM}) becomes 
\begin{eqnarray}
\label{Ener7EM}
E_{dipole}=-\frac{\sigma^{2}}{4\pi A}\frac{1-{\bf v}^{2}}{1+v^{2}}\left(1-{\bf v}^{2}\sin^{2}\theta\right)^{-1/2} \ ,
\end{eqnarray}
where, $A=\left|\bf A\right|$ and $\theta$ is the angle between the vectors ${\bf A}$ and $\bf v$. Notice that $0\leq\theta\leq\pi$. The energy (\ref{Ener7EM}) leads to an spontaneous torque on the dipole, as follows
\begin{eqnarray}
\label{TorqueEM}
\tau_{dipole}&=&-\frac{\partial E}{\partial\theta}=\frac{\sigma^{2}}{8\pi A}\frac{1-{\bf v}^{2}}{1+v^{2}}\frac{{\bf v}^{2}\sin(2\theta)}{\left(1-{\bf v}^{2}\sin^{2}\theta\right)^{3/2}}\cr\cr
&\cong&\frac{\sigma^{2}{\bf v}^{2}}{8\pi A}\sin(2\theta) \ .
\end{eqnarray}

Equation (\ref{TorqueEM}) shows that the torque on a dipole is an exclusive effect due to the Lorentz symmetry breaking. For $\theta=0,\pi/2,\pi$ the torque vanishes. When $\theta=\pi/4,$ the torque exhibits a maximum value. 

It is worthy mentioning that the results above can be naturally generalized for an arbitrary number of spacial dimensions, as in \cite{BaroneHidalgo1,BaroneHidalgo2}.

\section{A steady current line and a point-like charge}
\label{IV}

In this Section we study the interaction energy between a steady line current and a point-like stationary charge, the second model studied in this paper. The steady current line shall be taken to flow parallel to the $z$-axis, along the straight line located at ${\bf A}=(A^{1},A^{2},0)$. The electric charge is placed at position ${\bf a}_{(2)}$. The hole external source of the system is given by
\begin{eqnarray}
\label{corre3Em}
J_{\mu}^{II}\left(x\right)=\sigma_{1}\eta^{3}_{\ \mu}\delta^{2}\left({\bf x}_{\perp}-{\bf A}\right)+\sigma_{2}\eta^{0}_{\ \mu}\delta^{3}\left({\bf x}-{\bf a}_{(2)}\right) \ .
\end{eqnarray}
where ${\bf x}_{\perp}=(x^{1},x^{2},0)$, is the position vector perpendicular to the straight line current. The parameters $\sigma_{1}$ and $\sigma_{2}$ stand for, respectively, to the current intensity and electric charge strength. 

Substituting (\ref{corre3Em}) into (\ref{zxc1}), discarding self-interacting terms and using the fact that $D^{30}(x,y)=D^{03}(x,y)$, which can be seen from (\ref{propEm}), one can show that
\begin{eqnarray}
\label{Ener8EM}
E^{II}=\frac{\sigma_{1}\sigma_{2}}{T}\int\int d^{4}x\ d^{4}y\ D^{30}(x,y)\cr
\delta^{2}\left({\bf x}_{\perp}-{\bf A}\right)\delta^{3}\left({\bf y}-{\bf a}_{(2)}\right) \ .
\end{eqnarray}

Substituting the propagator (\ref{propEm}) in (\ref{zxc1}), performing (in this order) the integrals $d^{2}{\bf x}_{\perp}$, $d^{3}{\bf y}$, $dx^{3}$, $dp^{3}$, $dx^{0}$, $dp^{0}$ and $dy^{0}$ and identifying the time interval $\int dy^{0}=T$, we have
\begin{eqnarray}
\label{Ener9EM}
E^{II}=-\sigma_{1}\sigma_{2}\left(\frac{v^{3}v^{0}}{1+v^{2}}\right)\int\frac{d^{2}{\bf p}_{\perp}}{(2\pi)^{2}}\frac{\exp(i{\bf p}_{\perp}\cdot{\bf a}_{\perp})}{[{\bf p}_{\perp}^2-({\bf v}_{\perp}\cdot{\bf p}_{\perp})^{2}]} ,
\end{eqnarray}
where we defined ${\bf p}_{\perp}=(p^{1},p^{2},0)$ and the distance between the charge and the line current ${\bf a}_{\perp}={\bf A}-{\bf a}_{(2)}=(A^{1}-a^{1}_{(2)},A^{2}-a^{2}_{(2)},0)$.

In order to solve the integral above we employ the same method used in the previous Section, with the transformations (\ref{mudan1EM}), (\ref{defq}) and (\ref{zxc3}) carried out only for the perpendicular components to the straight line current of the vector involved, 
\begin{eqnarray}
\label{Ener10EM}
E^{II}=-\frac{\sigma_{1}\sigma_{2}}{\sqrt{1-{\bf v}_{\perp}^{2}}}\left(\frac{v^{3}v^{0}}{1+v^{2}}\right)\int\frac{d^{2}{\bf q}_{\perp}}{(2\pi)^{2}}\frac{\exp(i{\bf q}_{\perp}\cdot{\bf b}_{\perp})}{{\bf q}_{\perp}^2} .\cr
\ 
\end{eqnarray}

The integral in (\ref{Ener10EM}) is divergent. In order to circumvent this problem we proceed as in reference \cite{BaroneHidalgo1}, introducing a mass regulator parameter, as follows
\begin{eqnarray}
\label{Ener11EM}
E^{II}=-\frac{\sigma_{1}\sigma_{2}}{\sqrt{1-{\bf v}_{\perp}^{2}}}\left(\frac{v^{3}v^{0}}{1+v^{2}}\right)\cr
\lim_{m\rightarrow0}\int\frac{d^{2}{\bf q}_{\perp}}{(2\pi)^{2}}\frac{\exp(i{\bf q}_{\perp}\cdot{\bf b}_{\perp})}{{\bf q}_{\perp}^2+m^{2}} \ .
\end{eqnarray}
using the fact that \cite{BaroneHidalgo1}
\begin{eqnarray}
\label{int4EM}
\int\frac{d^{2}{\bf q}_{\perp}}{(2\pi)^{2}}\frac{\exp(i{\bf q}_{\perp}\cdot{\bf b}_{\perp})}{{\bf q}_{\perp}^2+m^{2}}=\frac{1}{2\pi}K_{0}(mb_{\perp}) \ ,
\end{eqnarray}
we write
\begin{eqnarray}
\label{Ener12EM}
E^{II}=-\frac{\sigma_{1}\sigma_{2}}{2\pi\sqrt{1-{\bf v}_{\perp}^{2}}}\left(\frac{v^{3}v^{0}}{1+v^{2}}\right)\lim_{m\rightarrow0}[K_{0}(mb_{\perp})] \ ,
\end{eqnarray}
where, $K_{0}(mb_{\perp})$ stands for the K-Bessel function and $b_{\perp}=\left|{\bf b}_{\perp}\right|$.

Now, we use the fact that \cite{Arfken} $K_{0}(mb_{\perp})\stackrel{m\rightarrow0}{\rightarrow}-\ln\left(mb_{\perp}/2\right)-\gamma$,
where $\gamma$ is the Euler constant, in order do handle the expression (\ref{Ener12EM}) as follows
\begin{eqnarray}
\label{Ener13EM}
E^{II}&=&\frac{\sigma_{1}\sigma_{2}}{2\pi\sqrt{1-{\bf v}_{\perp}^{2}}}\left(\frac{v^{3}v^{0}}{1+v^{2}}\right)\lim_{m\rightarrow0}\left[\ln\left(\frac{mb_{\perp}}{2}\right)+\gamma\right]\nonumber\\
&=&\frac{\sigma_{1}\sigma_{2}}{2\pi\sqrt{1-{\bf v}_{\perp}^{2}}}\left(\frac{v^{3}v^{0}}{1+v^{2}}\right)\lim_{m\rightarrow0}\biggl[\ln\biggl(\frac{mb_{\perp}}{2}\biggr)+\gamma\nonumber\\
&\ &+\ln(ma_{0})-\ln(ma_{0})\biggr]\nonumber\\
&=&\frac{\sigma_{1}\sigma_{2}}{2\pi\sqrt{1-{\bf v}_{\perp}^{2}}}\left(\frac{v^{3}v^{0}}{1+v^{2}}\right)\biggl[\ln\left(\frac{b_{\perp}}{a_{0}}\right)\nonumber\\
&\ &\left[\gamma-\ln2+\lim_{m\rightarrow0}\ln(ma_{0})\right]\biggr]\ .
\end{eqnarray}
where, in the third line, we added and subtract the quantity $\ln(ma_{0})$, where $a_{0}$ is a arbitrary length-dimensional constant. We can neglect the last line of Eq. (\ref{Ener13EM}) once it does not depend on the distance $a_{\perp}$ and, therefore, does not contribute to the force between line current and the point charge. Defining ${\hat I}$ as the unit vector along the straight line current and noticing that $v^{3}$ is the projection of the vector ${\bf v}$ along the straight line current, one can identify $v^{3}={\bf v}\cdot{\hat I}$. So, the energy can be written as
\begin{eqnarray}
\label{Ener14EM}
E^{II}=\frac{\sigma_{1}\sigma_{2}}{2\pi}\frac{{\bf v}\cdot{\hat I}}{\sqrt{1-{\bf v}_{\perp}^{2}}}\left(\frac{v^{0}}{1+v^{2}}\right)\ln\left(\frac{b_{\perp}}{a_{0}}\right) \ ,
\end{eqnarray}
where
\begin{eqnarray}
\label{modb1EM}
b_{\perp}=\sqrt{{\bf a}_{\perp}^{2}+\frac{({\bf v}_{\perp}\cdot{\bf a}_{\perp})^{2}}{1-{\bf v}_{\perp}^{2}}} \ .
\end{eqnarray}

At this point, some comments are in order. The interaction energy (\ref{Ener14EM}) is an effect due solely to the Lorentz-symmetry breaking and has no counterpart in Maxwell theory, where a point-like charge does not interact with a steady line current. This fact can be seen by noticing that in the limit $v^{\mu}=0$, the energy (\ref{Ener14EM}) vanishes. The energy (\ref{Ener14EM}) is proportional to the electric charge $\sigma_{2}$ as well as to projection of the Lorentz-symmetry breaking vector ${\bf v}$ along the current line, which is proportional to $\sigma_{1}{\bf v}\cdot{\hat I}$. If the current line flows perpendicular to ${\bf v}$, there is no interaction. If $v^{\mu}$ is a space-like vector and $v^{0}=0$ the energy (\ref{Ener14EM}) vanishes too. The spatial dependence of the force between the charge and the steady current falls down as the distance increases with an inverse law, in first order approximation. The exact expression for the force is
\begin{eqnarray}
{\bf F}^{II}=-\nabla_{{\bf a}_{\perp}}E^{II}=\frac{\sigma_{1}\sigma_{2}}{2\pi}\frac{{\bf v}\cdot{\hat I}}{\sqrt{1-{\bf v}_{\perp}^{2}}}\left(\frac{v^{0}}{1+v^{2}}\right)\cr
\Biggl[{\bf a}_{\perp}^{2}+\frac{({\bf v}_{\perp}\cdot{\bf a}_{\perp})^{2}}{1-{\bf v}_{\perp}^{2}}\Biggr]^{-1}\Biggl[{\bf a}_{\perp}+\frac{({\bf v}_{\perp}\cdot{\bf a}_{\perp})}{1-{\bf v}_{\perp}^{2}}{\bf v}_{\perp}\Biggr]
\end{eqnarray}

As a final comment, we point out that if we fix the point-charge, the energy (\ref{Ener14EM}) leads to a torque on the steady current, which can be calculated denoting by $\theta$ the angle between ${\bf v}_{\perp}$ and ${\bf a}_{\perp}$, as follows  
\begin{eqnarray}
\label{torquelc}
\tau^{II}&=&-\frac{\partial E^{II}}{\partial\theta}\cr
&=&-\frac{\sigma_{1}\sigma_{2}}{4\pi}\frac{{\bf v}\cdot{\hat I}}{\sqrt{1-{\bf v}_{\perp}^{2}}}\left(\frac{v^{0}}{1+v^{2}}\right)\frac{{\bf v}_{\perp}^{2}\sin(2\theta)}{1-{\bf v}_{\perp}^{2}\sin^{2}(\theta)}\ .
\end{eqnarray}
If $\theta=0,\pi/2,\pi$ the torque (\ref{torquelc}) vanishes.

\section{Dirac strings}
\label{V}

In this Section we consider the system composed by a point-like charge placed at position ${\bf a}$ and a Dirac string. This system is described by the external source
\begin{eqnarray}
\label{Dcurrent1}
J^{\mu,III}\left(x\right)=J_{(D)}^{\mu}\left(x\right)+\sigma\eta^{0\mu}\delta^{3}({\bf x}-{\bf a}) \ ,
\end{eqnarray}
where $J_{(D)}^{\mu}\left(x\right)$ stands for the external field source produced by the Dirac string and second term on the right hand side, the source produced by the point-like charge.

Now, we choose a coordinate system where the Dirac string lies along the $z$-axis with internal magnetic flux $\Phi$. Its corresponding source is given by \cite{FernandaDissertacao,AndersonDissertacao}
\begin{equation}
\label{Dircurr2}
J_{(D)}^{\mu}({\bf x})=i\Phi(2\pi)^{2}\int\frac{d^{4}p}{(2\pi)^{4}}\delta(p^{0})\delta(p^{3})\varepsilon^{0\mu}_{\ \ \nu3}\ p^{\nu}e^{-ipx}\ .
\end{equation}
If $\Phi>0$ we have the internal magnetic field along $\hat z$. For $\Phi<0$, the internal magnetic field points in the opposite direction. 

From now on, in this Section, the sub-index $\perp$ means we are taking just the components of a given vector perpendicular to the string. For instance, ${\bf p}_{\perp}=(p_{x},p_{y},0)$ is the momentum perpendicular do the string.

Substituting (\ref{Dircurr2}) into (\ref{Dcurrent1}) and taking this result into (\ref{zxc1}), using the Fourier representation (\ref{propEm}) for the propagator  $D^{\mu\nu}(x,y)$, discarding self interacting terms, which do not contribute to the force between the string and the charge (the self interacting terms are proportional to $\sigma^2$ or $\Phi^2$), and following similar steps employed in the previous sections, we can show that
\begin{eqnarray}
\label{EnerD15EM}
E^{III}=-i\sigma\Phi\left(\frac{v^{0}}{1+v^{2}}\right)\int\frac{d^{2}{\bf p}_{\perp}}{(2\pi)^{2}}\frac{\left[\hat{z}\cdot\left({\bf p}_{\perp}\times{\bf v}_{\perp}\right)\right]}{[{\bf p}_{\perp}^2-({\bf v}_{\perp}\cdot{\bf p}_{\perp})^{2}]}\cr\cr
\exp\left(i{\bf p}_{\perp}\cdot{\bf a}_{\perp}\right) \ .
\end{eqnarray}

Now, we make the change of integration variables (\ref{defq}), (\ref{mudan6EM}), (\ref{zxc2}), (\ref{zxc3}) and (\ref{mudan3EM}) just for the $\perp$ coordinates in order to rewrite expression (\ref{EnerD15EM}) in the form
\begin{eqnarray}
\label{EnerD16EM}
E^{III}=-\frac{\sigma\Phi}{\sqrt{1-{\bf v}^{2}_{\perp}}}\left(\frac{v^{0}}{1+v^{2}}\right)\left[\hat{z}\cdot\left({\bf \nabla}_{{\bf b}_{\perp}}\times{\bf v}_{\perp}\right)\right]\cr\cr
\lim_{m\rightarrow0}\int\frac{d^{2}{\bf q}_{\perp}}{(2\pi)^{2}}\frac{\exp\left(i{\bf q}_{\perp}\cdot{\bf b}_{\perp}\right)}{{\bf q}^{2}_{\perp}+m^{2}} \ ,
\end{eqnarray}
where we inserted a mass-dimensional regulator parameter $m$ in order to eliminate divergences and defined the differential operator
\begin{eqnarray}
\label{exchange}
{\bf\nabla}_{{\bf b}_{\perp}}=\left(\frac{\partial}{\partial b^{1}},\frac{\partial}{\partial b^{2}}\right) \ ,
\end{eqnarray}

Substituting (\ref{int4EM}) in (\ref{EnerD16EM}) and performing some simple calculations we can show that
\begin{eqnarray}
\label{EnerD17EM}
E^{III}=\lim_ {m\rightarrow0}\frac{\sigma\Phi}{2\pi\sqrt{1-{\bf v}^{2}_{\perp}}}\left(\frac{v^{0}}{1+v^{2}}\right)\left[\frac{mK_{1}\left(m{\bf b}_{\perp}\right)}{|{\bf b}_{\perp}|}\right]\cr\cr
\left[\hat{z}\cdot\left({\bf b}_{\perp}\times{\bf v}_{\perp}\right)\right] \ .
\end{eqnarray}

We now take the limit $m\rightarrow0$, use definition (\ref{zxc3}) and the fact that $\Phi{\hat z}=|\Phi|{\hat B}_{int}$, where ${\hat B}_{int}$ is the unit vector pointing in the internal magnetic field direction, what leads to  
\begin{eqnarray}
\label{EnerD19EM}
E^{III}=\frac{\sigma|\Phi|}{2\pi\sqrt{1-{\bf v}^{2}_{\perp}}}\left(\frac{v^{0}}{1+v^{2}}\right)\left[{\bf a}_{\perp}^{2}+\frac{({\bf v}_{\perp}\cdot{\bf a}_{\perp})^{2}}{1-{\bf v}_{\perp}^{2}}\right]^{-1}\cr\cr\left[{\hat B}_{int}\cdot\left({\bf a}_{\perp}\times{\bf v}_{\perp}\right)\right] .
\end{eqnarray}

We notice that (\ref{EnerD19EM}) is also an exclusive effect due to the Lorentz-symmetry breaking. In the limit $v^{\mu}\to0$ the energy (\ref{EnerD19EM}) vanishes.

The energy (\ref{EnerD19EM}) leads a force between the Dirac string and the charge as well as a torque on the string, by fixing the charge position.

The fourth example is given by two parallel Dirac strings placed a distance $a$ apart. We take a coordinate system where the first string lies along the $z$ axis, with internal magnetic flux $\Phi_{1}$ and a second one, with $\Phi_{2}$ lying along the line ${\bf a}=(a^{1},a^{2})$. The corresponding external source is given by 
\begin{eqnarray}
\label{duasDcurrent1}
J_{\mu}^{IV}\left({\bf x}\right)=J_{\mu(D,1)}\left({\bf x}\right)+J_{\mu(D,2)}\left({\bf x}\right) \ ,
\end{eqnarray}
where $J_{(D,1)}^{\mu}\left({\bf x}\right)$ has the same form of (\ref{Dircurr2}), with $\Phi$ replaced by $\Phi_{1}$, and
\begin{equation}
J_{(D,2)}^{\mu}\left({\bf x}\right)=i\Phi_{2}\int\frac{d^{4}p}{(2\pi)^{2}}\delta(p^{0})\delta(p^{3})\varepsilon^{0\mu}_{\ \ \nu3}\ p^{\nu}e^{-ipx}e^{-i{\bf p}_{\perp}\cdot{\bf a}}
\end{equation}

Proceeding as in the previous cases, we can show that the interaction energy between the two Dirac string is given by
\begin{eqnarray}
\label{EnerD7EM}
E^{IV}=\frac{L{\Phi}_{1}{\Phi}_{2}}{\sqrt{1-{\bf v}^{2}_{\perp}}}\Biggl[-\Biggl(1+\frac{{\bf v}_{\perp}^{2}}{1+v^{2}}\Biggr)\int\frac{d^{2}{\bf q}_{\perp}}{(2\pi)^{2}}e^{i{\bf q}_{\perp}\cdot{\bf b}_{\perp}}\cr\cr
+\Biggl(\frac{1}{1+v^{2}}-\frac{1}{1-{\bf v}_{\perp}^{2}}\Biggr)\int\frac{d^{2}{\bf q}_{\perp}}{(2\pi)^{2}}\frac{\left({\bf v}_{\perp}\cdot{\bf q}_{\perp}\right)^{2}}{{\bf q}^{2}_{\perp}}e^{i{\bf q}_{\perp}\cdot{\bf b}_{\perp}}\Biggr] .
\end{eqnarray}
where we identified the Dirac string length $L=\int dx^{3}$.

The first term inside the brackets of (\ref{EnerD7EM}) is the Dirac delta function $\delta^{2}({\bf b}_{\perp})$ and provided that  ${\bf a}_{\perp}$ is nonzero (and, so, ${\bf b}_{\perp}$ is also nonzero) this term vanishes. Using definition (\ref{exchange}) and inserting a mass parameter, as we have made in (\ref{Ener11EM}), we have
\begin{eqnarray}
\label{EnerD8EM}
E^{IV}=-\frac{L{\Phi}_{1}{\Phi}_{2}}{\sqrt{1-{\bf v}^{2}_{\perp}}}\Biggl(\frac{1}{1+v^{2}}-\frac{1}{1-{\bf v}_{\perp}^{2}}\Biggr)\cr\cr
\left({\bf v}_{\perp}\cdot{\bf \nabla}_{{\bf b}_{\perp}}\right)^{2}\lim_{m\rightarrow0}\int\frac{d^{2}{\bf q}_{\perp}}{(2\pi)^{2}}\frac{\exp\left(i{\bf q}_{\perp}\cdot{\bf b}_{\perp}\right)}{{\bf q}^{2}_{\perp}+m^{2}}\ .
\end{eqnarray}

Using result (\ref{int4EM}), taking the limit $m\to0$, acting with the differential operators and using (\ref{modb1EM}) we have
\begin{eqnarray}
\label{EnerD11EM}
{\cal E}^{IV}=\frac{E^{IV}}{L}=\frac{1}{2\pi}\frac{{\Phi}_{1}{\Phi}_{2}}{\sqrt{1-{\bf v}^{2}_{\perp}}}\Biggl(\frac{1}{1+v^{2}}-\frac{1}{1-{\bf v}_{\perp}^{2}}\Biggr)\cr
\left({\bf a}_{\perp}^{2}+\frac{({\bf v}_{\perp}\cdot{\bf a}_{\perp})^{2}}{1-{\bf v}_{\perp}^{2}}\right)^{-2}\cr\cr
\Biggl[{\bf v}_{\perp}^{2}{\bf a}_{\perp}^{2}-\Biggr(\frac{2-{\bf v}_{\perp}^{2}}{1-{\bf v}_{\perp}^{2}}\Biggr)({\bf v}_{\perp}\cdot{\bf a}_{\perp})^{2}\Biggr]\ .
\end{eqnarray}
where we defined the energy per unit of string length ${\cal E}$.

It can be seen that the energy given above vanishes in the limit $v^{\mu}=0$, where we do not have Lorentz-symmetry breaking. The interaction energy (\ref{EnerD11EM}) is an effect due solely to the Lorentz symmetry breaking and has no counterpart in Maxwell theory.

The fifth and last model we consider is given by a Dirac string and a steady line current, both parallel to each other. The corresponding external source is
\begin{equation}
J^{V}_{\mu}({\bf x})=\sigma\eta^{3}_{\ \mu}\delta^{2}\left({\bf x}_{\perp}-{\bf A}\right)+J_{(D)}^{\mu}\left({\bf x}\right)
\end{equation}
where $J_{(D)}^{\mu}\left({\bf x}\right)$ is given by (\ref{Dircurr2}).

Following similar steps employed previously we have
\begin{eqnarray}
\label{EnerD21EM}
{\cal E}^{V}=\frac{E^{V}}{L}=\frac{\sigma|\Phi|}{2\pi\sqrt{1-{\bf v}^{2}_{\perp}}}\left(\frac{{\bf v}\cdot{\hat I}}{1+v^{2}}\right)\cr\cr
\left[{\bf a}_{\perp}^{2}+\frac{({\bf v}_{\perp}\cdot{\bf a}_{\perp})^{2}}{1-{\bf v}_{\perp}^{2}}\right]^{-1}\left[{\hat B}_{int}\cdot\left({\bf a}_{\perp}\times{\bf v}_{\perp}\right)\right] \ .
\end{eqnarray}

We can note that (\ref{EnerD21EM}) also represents an exclusive effect of the Lorentz symmetry breaking.

\section{Conclusions and final remarks}
\label{conclusoes}

In this paper, the interaction between stationary electromagnetic field sources were investigated in a Lorentz-symmetry breaking scenario in $(3+1)$ dimensions. The Lorentz-symmetry breaking parameter has been chosen to be a vector one, in such a way that all the results obtained are exact. We observed effects with no counterpart in Maxwell electrodynamics and not explored in the literature up to now. We showed the emergence of an spontaneous torque on a classical electromagnetic dipole and an interaction between a steady straight line current and a point-like charge. Both phenomena are due solely to the Lorentz-symmetry breaking.

We also investigate some phenomena due to the presence of a Dirac string. We show that this string can interact with a point charge as well as with a straight line steady current. We also study the interaction between two parallel Dirac strings. These results suggest that in models which exhibit explicit Lorentz-symmetry breaking the Aharonov-Bohm effect can be modified in comparison with the results predicted by Maxwell electrodynamics. This subject deserves more investigations not only in the context of the model considered in this paper, but also for other kinds of models with Lorentz-symmetry breaking.

Finally, we would like to comment that taking into account the spin of the sources would be a natural follow-up to the investigation we have pursued in the present paper. If we consider the situation where the charge carriers (our sources) may have spin 1/2 or 1 (here, our sources are spinless), the expressions for the interacting energies and forces worked out here shall display a spin dependence; not only a spin-spin interaction between the sources come out, but also a non-trivial coupling between the source's spin and the background vector shows up. Then, the spin-dependence of the interaction energy and force may disclose a number of peculiar effects with no counterpart in the usual Maxwellian Electrodynamics. We shall soon be reporting on the results of these studies \cite{BBH}.

\begin{acknowledgments}
L.H.C. Borges thanks to FAPESP, under the process 2013/01231-6 for financial support. F.A. Barone and J.A. Helay\"el-Neto thank to CNPq for financial support.
The authors thank F.E. Barone for suggestions and for reading the paper. 
\end{acknowledgments}



\end{document}